\documentclass[onecolumn]{aastex63}

\usepackage{amsmath}
\usepackage{afterpage}
\usepackage{scalerel}
\bibpunct{(}{)}{;}{a}{}{,}
\usepackage{multirow}
\usepackage{graphics}
\usepackage{threeparttable}
\usepackage[varg]{txfonts}
\usepackage{xcolor}
\usepackage{bm}
\hypersetup{colorlinks,linkcolor={blue},citecolor={blue},urlcolor={blue}}

\begin{document}
\title{A Multi-Band Forced-Photometry Catalog in the ELAIS-S1 Field}

\author[0000-0002-4436-6923]{Fan Zou}
\affiliation{Department of Astronomy and Astrophysics, 525 Davey Lab, The Pennsylvania State University, University Park, PA 16802, USA}
\affiliation{Institute for Gravitation and the Cosmos, The Pennsylvania State University, University Park, PA 16802, USA}

\author[0000-0002-0167-2453]{W. N. Brandt}
\affiliation{Department of Astronomy and Astrophysics, 525 Davey Lab, The Pennsylvania State University, University Park, PA 16802, USA}
\affiliation{Institute for Gravitation and the Cosmos, The Pennsylvania State University, University Park, PA 16802, USA}
\affiliation{Department of Physics, 104 Davey Laboratory, The Pennsylvania State University, University Park, PA 16802, USA}

\author[0000-0002-3032-1783]{Mark Lacy}
\affiliation{National Radio Astronomy Observatory, 520 Edgemont Road, Charlottesville, VA 22903, USA}

\author[0000-0002-8577-2717]{Qingling Ni}
\affiliation{Department of Astronomy and Astrophysics, 525 Davey Lab, The Pennsylvania State University, University Park, PA 16802, USA}
\affiliation{Institute for Gravitation and the Cosmos, The Pennsylvania State University, University Park, PA 16802, USA}

\author[0000-0003-1991-370X]{Kristina Nyland}
\affiliation{National Research Council, resident at the U.S. Naval Research Laboratory, 4555 Overlook Ave. SW, Washington, DC 20375, USA}

\author[0000-0001-8835-7722]{Guang Yang}
\affiliation{Department of Physics and Astronomy, Texas A$\&$M University, College Station, TX, 77843-4242 USA}
\affiliation{George P. and Cynthia Woods Mitchell Institute for Fundamental Physics and Astronomy, Texas A$\&$M University, College Station, TX, 77843-4242 USA}

\author[0000-0002-8686-8737]{Franz E. Bauer}
\affiliation{Instituto de Astrof\'isica, Facultad de F\'isica, Pontificia Universidad Cat\'olica de Chile Av. Vicu\~na Mackenna 4860, 782-0436 Macul, Santiago, Chile}
\affiliation{National Radio Astronomy Observatory, Pete V. Domenici Array Science Center, P.O. Box O, Socorro, NM 87801, USA}
\affiliation{Space Science Institute, 4750 Walnut Street, Suite 205, Boulder, CO 80301, USA}

\author[0000-0002-2553-096X]{Giovanni Covone}
\affiliation{INAF - Osservatorio Astronomico di Capodimonte, Salita Moiariello 16, I-80131, Napoli, Italy}
\affiliation{Dipartimento di Fisica, Universit\`a di Napoli ``Federico II'', via Cinthia 9, 80126 Napoli, Italy}
\affiliation{INFN - Sezione di Napoli, via Cinthia 9, 80126 Napoli, Italy}

\author[0000-0002-0501-8256]{Aniello Grado}
\affiliation{INAF - Osservatorio Astronomico di Capodimonte, Salita Moiariello 16, I-80131, Napoli, Italy}

\author[0000-0003-0911-8884]{Nicola R. Napolitano}
\affiliation{INAF - Osservatorio Astronomico di Capodimonte, Salita Moiariello 16, I-80131, Napoli, Italy}

\author[0000-0003-4210-7693]{Maurizio Paolillo}
\affiliation{Dipartimento di Fisica, Universit\`a di Napoli ``Federico II'', via Cinthia 9, 80126 Napoli, Italy}
\affiliation{INFN - Sezione di Napoli, via Cinthia 9, 80126 Napoli, Italy}
\affiliation{INAF - Osservatorio Astronomico di Capodimonte, Salita Moiariello 16, I-80131, Napoli, Italy}

\author[0000-0002-3585-866X]{Mario Radovich}
\affiliation{INAF - Osservatorio Astronomico di Padova, vicolo Osservatorio, 5 I-35122 Padova, Italy }

\author[0000-0002-6427-7039]{Marilena Spavone}
\affiliation{INAF - Osservatorio Astronomico di Capodimonte, Salita Moiariello 16, I-80131, Napoli, Italy}

\author[0000-0002-6748-0577]{Mattia Vaccari}
\affiliation{Inter-university Institute for Data Intensive Astronomy, Department of Physics and Astronomy, University of the Western Cape, Robert Sobukwe Road, 7535 Bellville, Cape Town, South Africa}
\affiliation{INAF - Istituto di Radioastronomia, via Gobetti 101, 40129 Bologna, Italy}

\email{E-mail: fuz64@psu.edu}


\begin{abstract}
The ELAIS-S1 field will be an LSST Deep Drilling field, and it also has extensive multiwavelength coverage. To improve the utility of the existing data, we use \textit{The Tractor} to perform forced-photometry measurements in this field. We compile data in 16 bands from the DeepDrill, VIDEO, DES, ESIS, and VOICE surveys. Using a priori information from the high-resolution fiducial images in VIDEO, we model the images in other bands and generate a forced-photometry catalog. This technique enables consistency throughout different surveys, deblends sources from low-resolution images, extends photometric measurements to a fainter magnitude regime, and improves photometric-redshift estimates. Our catalog contains over 0.8 million sources covering a $3.4~\mathrm{deg^2}$ area in the VIDEO footprint and is available at\dataset[10.5281/zenodo.4540178]{\doi{10.5281/zenodo.4540178}}.
\end{abstract}

\section*{}
\begin{figure*}
\resizebox{\hsize}{!}{
\includegraphics{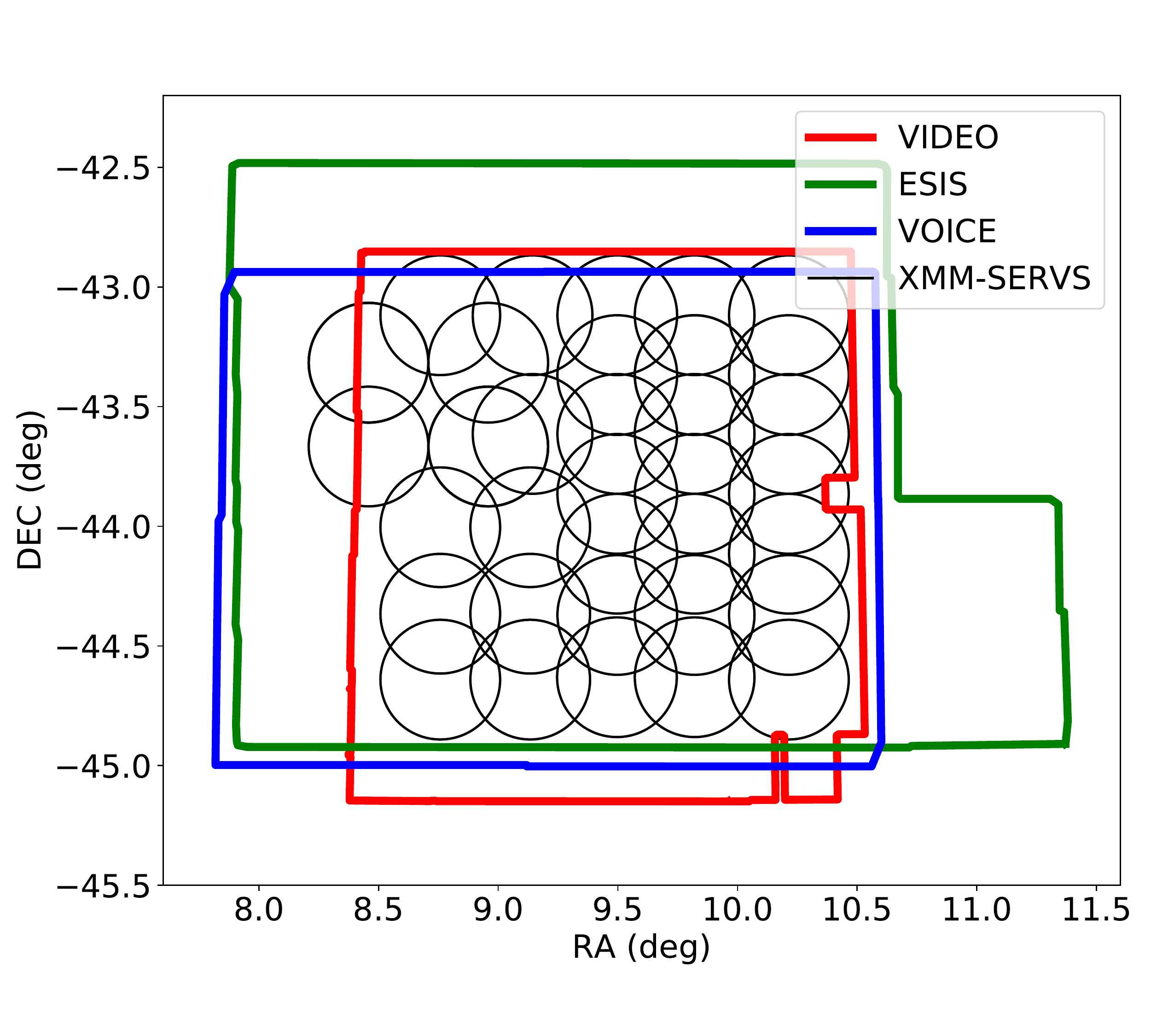}
}
\caption{The boundaries of multiwavelength survey regions in ELAIS-S1. The black circles represent the \textit{XMM-Newton} pointings composing the XMM-SERVS survey in this field. The \textit{Spitzer} DeepDrill and DES surveys cover the whole field, and thus their coverages are not displayed. Our cataloged $3.4~\mathrm{deg^2}$ region is the VIDEO survey region.}
\label{Fig_surveyedges}
\end{figure*}

To aid counterpart characterization for the XMM-SERVS survey\footnote{\url{http://personal.psu.edu/wnb3/xmmservs/xmmservs.html}} (\citealt{Chen18}; Ni et al., in preparation) and help prepare for the upcoming Deep Drilling Fields (e.g., \citealt{Brandt18}) of the Vera C. Rubin Observatory Legacy Survey of Space and Time (LSST), we generate a multi-band photometric catalog in the European Large-Area ISO Survey-S1 (ELAIS-S1) field from the \textit{Spitzer} DeepDrill (3.6 and $4.5~\mathrm{\mu m}$; \citealt{Lacy21})\footnote{\url{https://irsa.ipac.caltech.edu/data/SPITZER/DeepDrill/}}, VIDEO ($ZYJHK_s$; \citealt{Jarvis13})\footnote{\url{http://horus.roe.ac.uk/vsa/}}, DES DR2 ($grizy$; \citealt{Abbott21})\footnote{\url{https://des.ncsa.illinois.edu/releases/dr2}}, ESIS ($BVR$; \citealt{Berta06, Vaccari16})\footnote{\url{http://www/mattiavaccari.net/esis/}}, and VOICE ($u$; \citealt{Vaccari16})\footnote{\url{http://www.mattiavaccari.net/voice/}} surveys using \textit{The Tractor} \citep{Lang16} forced-photometry approach (see below). The coverages of these surveys are displayed in Fig.~\ref{Fig_surveyedges}. Compared to the conventional single-band photometric measurements, this forced-photometry technique can distinguish blended sources in low-resolution surveys (e.g., DeepDrill) and enables much more reliable photometric-redshift measurements (e.g., \citealt{Lang14, Nyland17, Wold19}).\par
We follow the same approach to make \textit{The Tractor} photometric measurements as \citet{Nyland17} for VIDEO-detected sources in the full VIDEO area. We only briefly outline the procedures here, and interested readers can refer to \citet{Nyland17} for more details. For each source, we adopt the band with the longest wavelength among the VIDEO bands in which the source is detected as the fiducial band and then obtain the source position and intrinsic surface-brightness profile (point-like, deVaucouleurs, or exponential profile) from the corresponding VIDEO catalog and image. We fix this prior position and intrinsic surface-brightness profile, only allowing the fluxes to vary, and fit the images in other bands, where the intrinsic profile is convolved with the corresponding PSFs to match the observed profiles. This procedure returns forced photometry. The photometric errors are derived using the equations in the Appendix of \citet{Nyland17}. The astrometric errors and image offsets of our data are smaller than the sizes of single pixels ($0.2''-0.3''$), and thus do not considerably downgrade our photometric accuracy \citep{Nyland17}. We flag possibly saturated sources as those with magnitudes smaller than 13.5 ($4.5~\mathrm{\mu m}$), 14.0 ($3.6~\mathrm{\mu m}$), 14.0 ($K_s$), 14.2 ($H$), 14.5 ($J$), 13.6 ($Y$), 14.0 ($Z$), 14.5 ($R$), 14.2 ($V$), and 14.6 ($B$). These thresholds are similar to those in \citet{Nyland17}. After comparing \textit{The Tractor} photometry and original photometry for DES, we find that there are some significant outliers. These outliers are not a result of \textit{The Tractor} fitting; instead, their original DES $z$- or $y$-band magnitudes also show significant deviations compared to their original VIDEO $Z$- or $Y$-band magnitudes, indicating that their DES images may be problematic (e.g., saturated), and thus these sources are flagged.\footnote{See, e.g., part 3 in \url{https://herschel-vos.phys.sussex.ac.uk/vo-data/dmu6/dmu6_v_ELAIS-S1/help_elais-s1_checks.html}, and we follow the same approach to define outliers.}\par

We release our catalog online.\footnote{\url{https://doi.org/10.5281/zenodo.4540178}} It contains fiducial bands, intrinsic surface-brightness profile, AB magnitudes, magnitude errors, flags, reduced $\chi^2$ values, and nearest-neighbor angular separations. The separations can be used as a basic indication for possible blending issues in DeepDrill images. A ``\texttt{readme.txt}'' file along with the catalog provides detailed explanations of the catalog columns. Our released data also include several diagnostic figures to characterize the reliability of our catalog: Fig.~S1 compares the forced photometry and the original photometry, with flagged sources displayed explicitly. Fig.~S2 plots \textit{The Tractor} magnitude errors versus magnitudes for all sources, based on which the nominal $5~\sigma$ magnitude limits are estimated. The limits are estimated to be 24.3 ($4.5~\mathrm{\mu m}$), 24.5 ($3.6~\mathrm{\mu m}$), 24.6 ($K_s$), 25.1 ($H$), 25.7 ($J$), 26.0 ($Y$), 26.2 ($Z$), 22.7 ($y$), 23.9 ($z$), 24.5 ($i$), 25.2 ($r$), 25.5 ($g$), 25.8 ($R$), 25.9 ($V$), 26.5 ($B$), and 25.9 ($u$). Fig.~S3 displays the magnitude distributions for both the original catalogs and our catalog after applying a $2~\sigma$ cut and demonstrates that we are able to place photometric constraints on a considerable number of faint sources not formally detected in single-band images.\par
Our catalog greatly improves the photometric-redshift ($z_\mathrm{phot}$) results, which was a primary motivation for our work. We use EAZY \citep{Brammer08} to estimate $z_\mathrm{phot}$, following an iterative method of magnitude zero-point correction \citep{Yang14}, and then compare $z_\mathrm{phot}$ values and spectroscopic redshifts ($z_\mathrm{spec}$) for $\sim10000$ sources with $z_\mathrm{spec}$ measurements (typically with $R$ magnitudes $\approx19-23$), where the $z_\mathrm{spec}$ compilation is described in Ni et al. (in preparation). Using publicly available SERVS, VIDEO, and DES cross-matched catalogs, the normalized median absolute deviation ($\sigma_\mathrm{NMAD}$) and the outlier fraction ($f_\mathrm{outlier}$; defined as $|z_\mathrm{phot}-z_\mathrm{spec}|/(1+z_\mathrm{spec})>0.15$) are 0.071 and 17.4\%, respectively. After utilizing our catalog, the values drop to $\sigma_\mathrm{NMAD}=0.032$ and $f_\mathrm{outlier}=4.1\%$. Interested readers can refer to Ni et al. (in preparation) for more details on the $z_\mathrm{phot}$ measurements as well as the official release of the corresponding $z_\mathrm{phot}$ catalog.\par
The forced-photometry catalogs in the other parts of XMM-SERVS are or will be publicly available as well; see \citet{Nyland17} and Nyland et al. (in preparation).
\bigbreak
\textit{Acknowledgements}. We acknowledge support from NASA grant 80NSSC19K0961. Basic research in radio astronomy at the U.S. Naval Research Laboratory is supported by 6.1 Base Funding. FEB acknowledges support from ANID-Chile Basal AFB-170002, FONDECYT Regular 1200495 and 1190818, and Millennium Science Initiative – ICN12\_009. MV acknowledges support from the Italian Ministry of Foreign Affairs and International Cooperation (MAECI Grant Number ZA18GR02) and the South African Department of Science and Technology's National Research Foundation (DST-NRF Grant Number 113121) as part of the ISARP RADIOSKY2020 Joint Research Scheme.

\newpage

{}

\appendix
Here are some supplementary notes:
\begin{itemize}
\item{Fig.~S1. The slightly tilted trends in the DeepDrill panels may be due to calibration differences for extended sources and also exist in other work (e.g., \citealt{Nyland17}). The small ``antenna-shaped'' structures in the VIDEO panels, where the original magnitudes are higher than \textit{The Tractor} magnitudes for some faint sources, are mainly composed of sources around bright stars, and we attribute much of these source fluxes to the backgrounds contributed by the adjacent stars. The degree to which the plots appeal to ``funnel'' out in the faint regime depends on the detection significance thresholds of the original catalogs as well as the total number of plotted sources (or the plotting settings). For example, the original ESIS catalog has a relatively higher threshold and fewer sources, and thus the noise of the ESIS panels is less.}
\item{DES DR2 outliers. DES outliers are selected by comparing VIDEO $ZY$ and DES $zy$ magnitudes (see Footnote 7). Fig.~S1 further shows that \textit{The Tractor} magnitudes of these outliers also deviate from the corresponding DES magnitudes in all the DES bands, and thus we suggest applying the DES outlier flags to all the DES bands. Although many of these outliers present clearly saturated images (i.e., with trapezoidal brightness profiles), the others do not show abnormal images. The actual causes for these outliers, however, are still unclear.}
\end{itemize}

\renewcommand\thefigure{S1}
\begin{figure*}[htbp]
\resizebox{\hsize}{!}{
\includegraphics{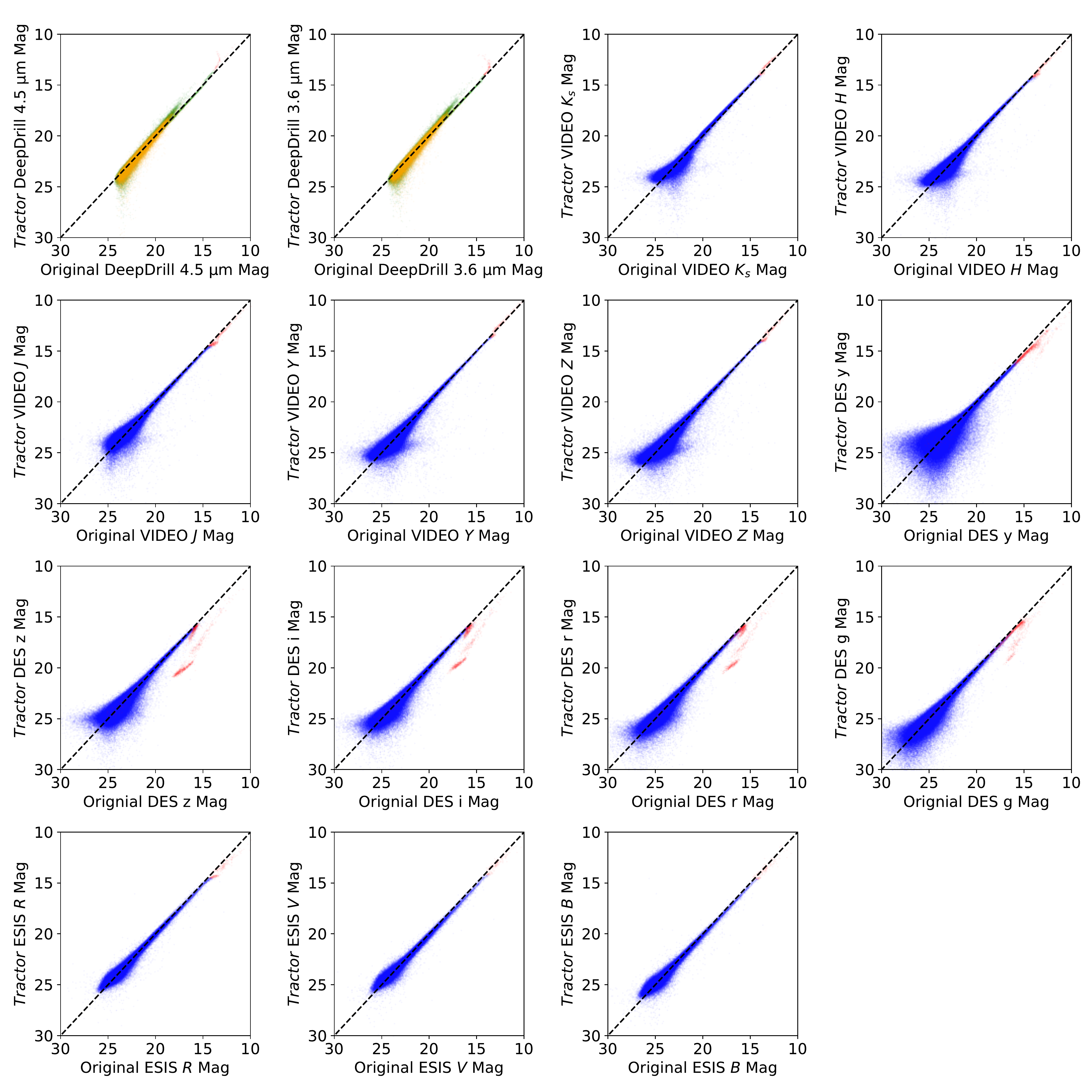}
}
\caption{Comparison between \textit{The Tractor} photometry and original AB photometry for all the bands except $u$ band, which only has available images but does not have an original catalog. The black dashed lines mark one-to-one relations. The red points are flagged sources. For DeepDrill, the orange points are possibly blended sources (i.e., nearest-neighbor separations $<3.8''$), and the green points are other sources, which are less likely to suffer from blending issues. The original magnitudes used for comparisons here are aperture-corrected magnitudes within an aperture of radius = $1.9''$ for DeepDrill, Petrosian magnitudes for VIDEO, and MAG\_AUTO magnitudes for DES and ESIS.}
\label{Fig_compare_inwout}
\end{figure*}

\renewcommand\thefigure{S2}
\begin{figure*}[htbp]
\resizebox{\hsize}{!}{
\includegraphics{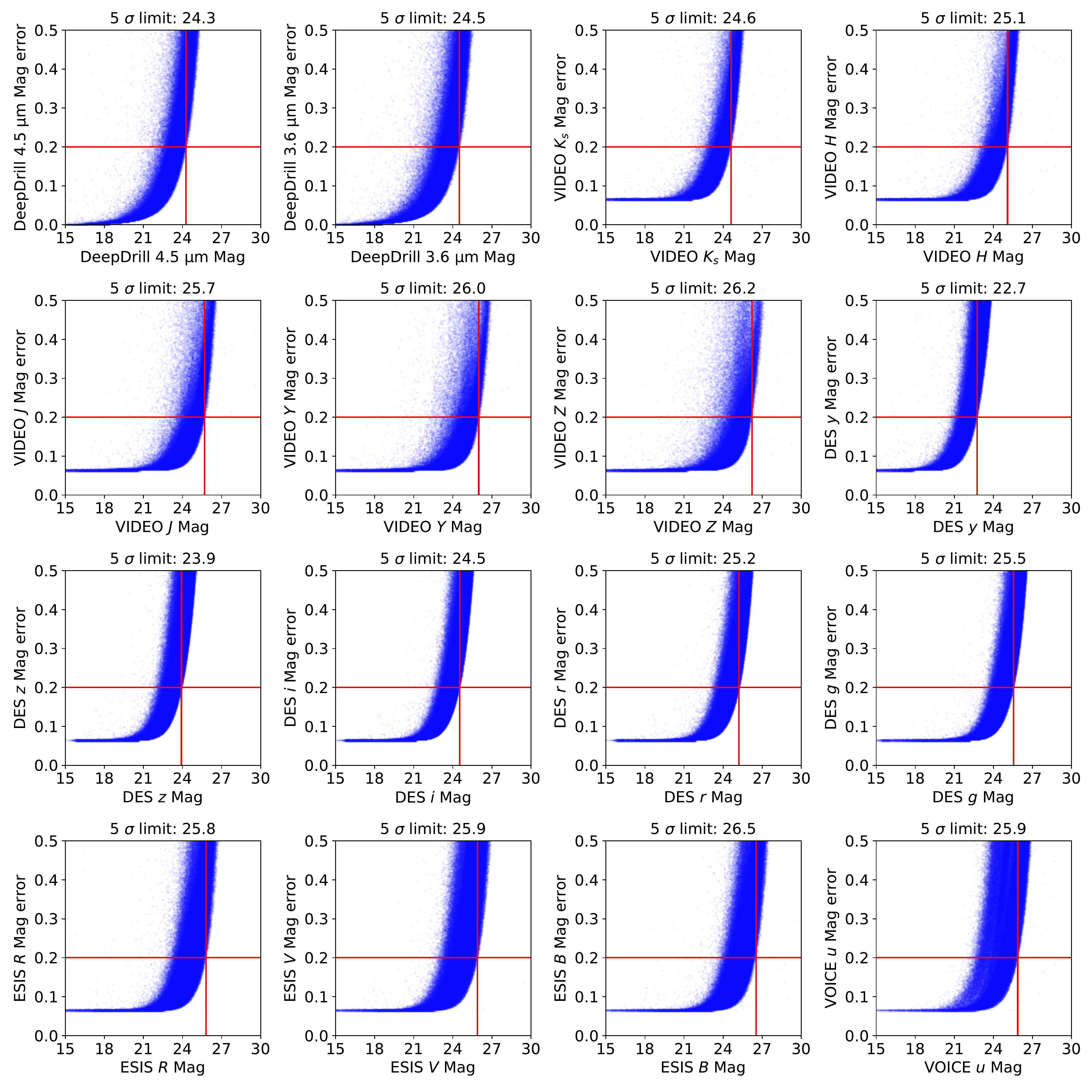}
}
\caption{\textit{The Tractor} magnitude errors versus AB magnitudes. The red lines mark the $5~\sigma$ limits, where the errors are 0.2, and the limits are displayed as the titles. Note that the actual limits vary among different regions in the field for non-uniform surveys (e.g., the $u$ band of the VOICE survey, whose exposure time varies across the whole field), and the limits quoted here only refer to the deepest regions.}
\label{Fig_errorlimit}
\end{figure*}

\renewcommand\thefigure{S3}
\begin{figure*}[htbp]
\resizebox{\hsize}{!}{
\includegraphics{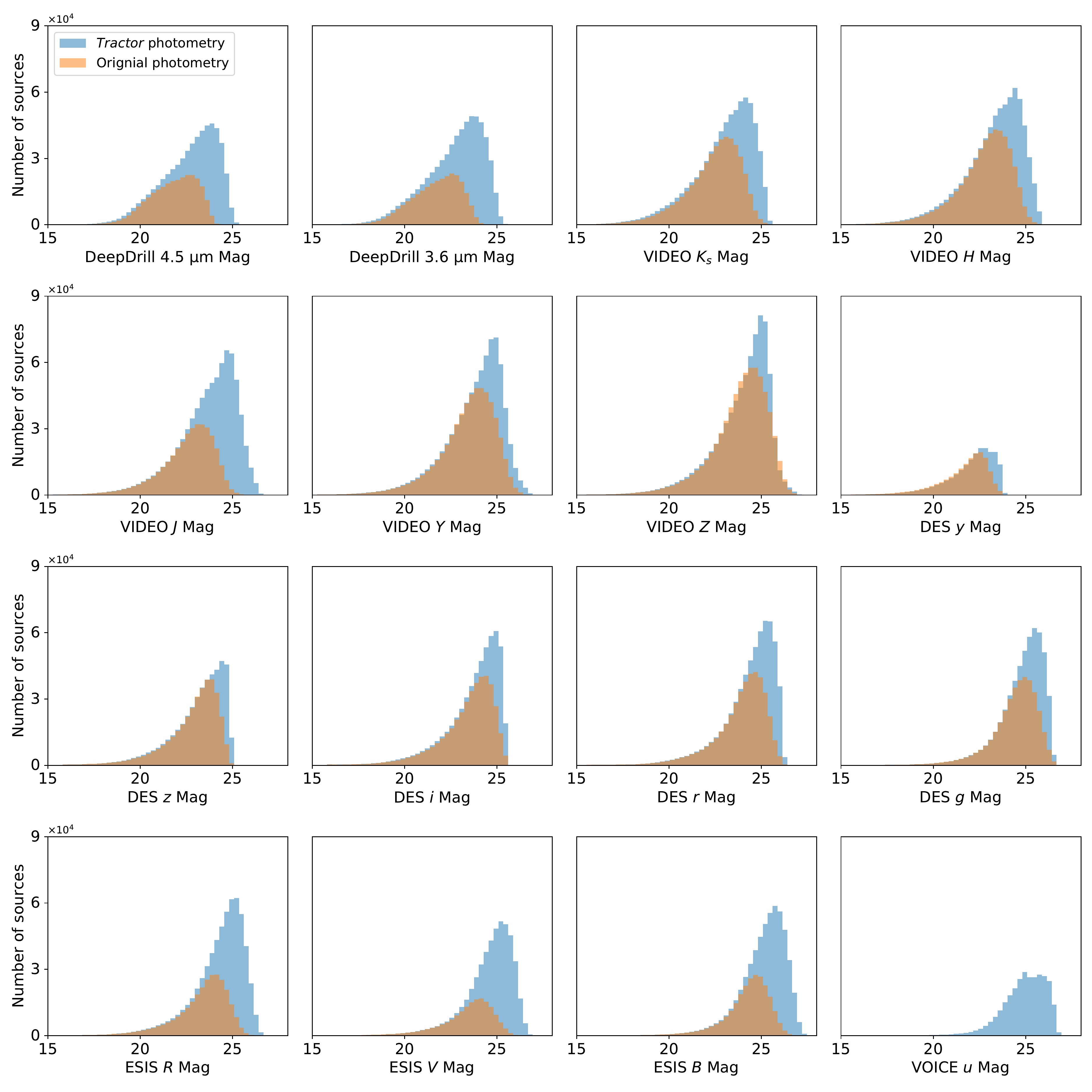}
}
\caption{Magnitude distributions for our catalog from \textit{The Tractor} and original catalogs in each band. Only sources with magnitude errors smaller than 0.54 are shown, where the threshold roughly corresponds to signal-to-noise ratios = 2 (e.g., see Eq.~4 in \citealt{Abbott21}). No original catalog is available for the VOICE $u$ band.}
\label{Fig_hist_compare}
\end{figure*}

\end{document}